\documentclass[]{spie}
\usepackage[]{graphicx}

\usepackage[colorlinks=true,linkcolor=blue,anchorcolor=blue,citecolor=blue,urlcolor=blue]{hyperref}

\title{Weak localization and weak anti-localization in topological insulators}

\author{Hai-Zhou Lu and Shun-Qing Shen
\skiplinehalf
Department of Physics, The University of Hong Kong, Pofulam Road, Hong Kong, China}

\begin{document}
\maketitle

\begin{abstract}

Weak localization and weak anti-localization are quantum interference effects in quantum transport in a disordered electron system. Weak anti-localization enhances the conductivity and weak localization suppresses the conductivity with decreasing temperature  at very low temperatures. A magnetic field can destroy the quantum interference effect, giving rise to a cusp-like positive and negative magnetoconductivity as the signatures of weak localization and weak anti-localization, respectively. These effects have been widely observed in topological insulators. In this article, we review recent progresses in both theory and experiment of weak (anti-)localization in topological insulators, where the quasiparticles are described as Dirac fermions. We predicted a crossover from weak anti-localization to weak localization if the massless Dirac fermions (such as the surface states of topological insulator) acquire a Dirac mass, which was confirmed experimentally. The bulk states in a topological insulator thin film can exhibit the weak localization effect, quite different from other system with strong spin-orbit interaction.
We compare the localization behaviors of Dirac fermions with conventional electron systems in the presence of disorders of different symmetries. Finally, we show that both the interaction and quantum interference are required to account for the experimentally observed temperature and magnetic field dependence of the conductivity at low temperatures.
\end{abstract}


\tableofcontents

\section{INTRODUCTION}
\label{sec:intro}

Topological insulators are gapped band insulators with topologically protected gapless modes surrounding their boundaries \cite{Moore10nat,Hasan10rmp,Qi11rmp,Shen12book}. The surface states of a three-dimensional topological insulator are composed of an odd number of two-dimensional gapless Dirac cones, which has a helical spin structure in momentum space\cite{Xia09natphys}, giving rise to a $\pi$ Berry phase when an electron moves adiabatically around the Fermi surface. The $\pi$ Berry phase can lead to the absence of backscattering\cite{Ando98jpsj},  and further delocalization of the surface electrons \cite{Bardarson07prl,Nomura07prl}.
Because of these properties, the topological insulators are believed to have better performance in future electronic devices, thus have attracted much interest in studying their transport properties\cite{Culcer12pe,Bardarson13rpp}.

In experiments, the delocalization of electrons was believed to be verified by a phenomenon named the weak anti-localization\cite{HLN80}. The effect stems from the $\pi$ Berry phase\cite{Shen04prb}, which induces a destructive quantum interference between time-reversed loops formed by scattering trajectories. The destructive interference can suppress backscattering of electrons, then the conductivity is enhanced with decreasing temperature because decoherence mechanisms are suppressed at low temperatures\cite{Suzuura02prl,McCann06prl}. A magnetic field can destroy the interference as well as the enhanced conductivity, so the signature of the weak anti-localization is a negative magnetoconductivity, which has been observed in a lot of topological insulator samples\cite{Checkelsky09prl,Peng10natmat,Chen10prl,Checkelsky11prl,He11prl,Kim11prb,Steinberg11prb}.

Having topologically protected surface states that cannot be localized is one of the alternative definitions of the topological insulators\cite{Moore10nat,Hasan10rmp}, and this is regarded as one of the merits of a topological insulator compared to ordinary metals. However, the metallic enhancement in conductivity expected to appear along with the negative magnetoconductivity is not observed at low temperatures. Instead, in most experiments, a logarithmic suppression of the conductivity with decreasing temperature is observed\cite{Wang11prb,Liu11prb,Chen11rc,Takagaki12prb,Chiu13prb,Roy13apl}, indicating a behavior of the weak localization, which was supposed to occur in ordinary disordered metals as a precursor of the Anderson localization\cite{Bergmann84PhysRep,Lee85rmp}, which was known as a transport paradox in topological insulators.

Here we review our recent efforts\cite{Lu11prl,Lu11prb,Shan12prb,Lu14prl} on the theoretical understanding to the weak localization and weak anti-localization effects in the transport experiments in topological insulators.
In Sec. \ref{sec:WL-WAL}, we give an introduction to the quantum diffusion regime, where the weak (anti-)localization happens. Then the experiments of weak anti-localization in topological insulators are reviewed. In Sec. \ref{sec:crossover}, we discuss the crossover between weak anti-localization and weak localization and the Berry phase argument. In Sec. \ref{fig:bulk}, we show why the bulk states in topological insulator can have weak localization. In Sec. \ref{sec:conventional}, we compare the Dirac fermions with conventional electrons, on their localization behaviors in the presence of three kinds of disorder scattering. In Sec. \ref{sec:interaction}, we will explain how the contradictory observations in the temperature and magnetic field dependence of the conductivity of topological insulators can be understood, by including both the quantum interference and electron-electron interactions for the disordered Dirac fermions.

\section{WEAK LOCALIZATION AND ANTI-LOCALIZATION}
\label{sec:WL-WAL}

\subsection{Quantum diffusion}
\begin{figure}[htbp]
\centering \includegraphics[width=0.6\textwidth]{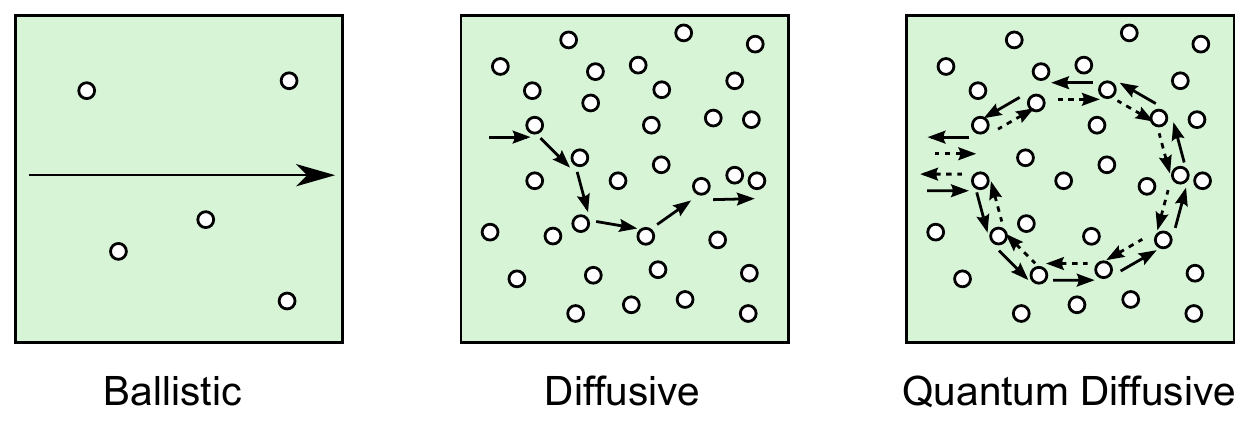}
\caption{Schematic illustration of different electronic transport regimes in solids. The open circles represent impurities and arrows mark the trajectories that electron travelled. }
\label{fig:transport}
\end{figure}

The electronic transport in solids can be classified by several characteristic lengths: (i) The mean free path $\ell$, which measures the average distance that an electron travels before its momentum is changed by elastic scattering from static scattering centers. (ii) The phase coherence length $\ell_\phi$, which measures the average distance an electron can maintain its phase coherence. $\ell_\phi$ is usually determined by inelastic scattering from electron-phonon coupling and interaction with other electrons. (iii) The sample size $L$.

If $\ell\gg L$, electrons can tunnel through the sample without being scattered. This is the ballistic transport regime. In the opposite limit $\ell\ll L$, electrons will suffer from scattering and diffuse through the sample, and this is the diffusive transport regime. In the diffusive regime, if $\ell_\phi \le \ell $, we call it the semiclassical diffusion, and this part gives the Drude conductivity. If $\ell_\phi\gg \ell$, electrons will maintain their phase coherence even after being scattered for many times, referring to as the quantum diffusive regime. In this regime, the quantum interference between time-reversed scattering loops (see Fig. \ref{fig:transport}) will give rise to a correction to the conductivity. The weak localization or weak anti-localization is an effect due to this correction in the conductivity in the quantum diffusive regime.

\subsection{Signatures of weak (anti-)localization}

\begin{figure}[htbp]
\centering \includegraphics[width=0.6\textwidth]{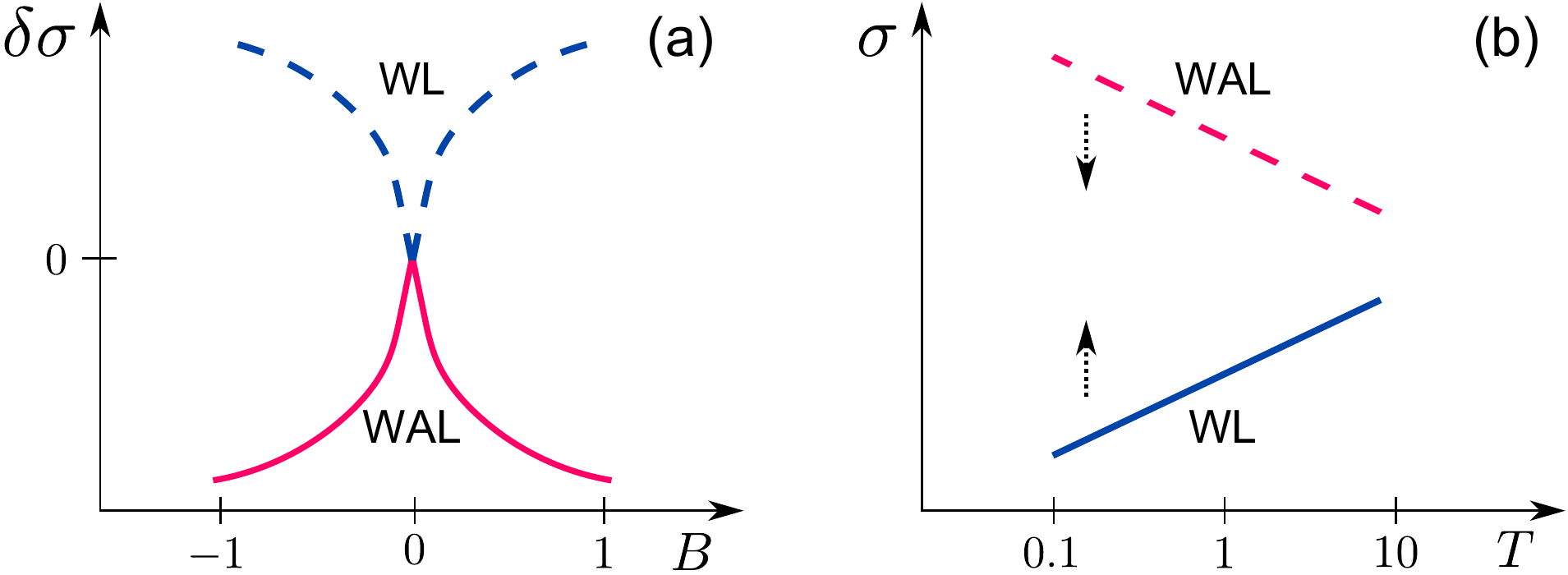}
\caption{The signatures of weak localization (WL) and weak anti-localization (WAL) in two dimensions in (a) magnetoconductivity [defined as $\delta \sigma\equiv \sigma(B)-\sigma(0)$] and (b) temperature dependence of the conductivity $\sigma$. $B$ is magnetic field and $T$ is temperature. Adapted from Ref. \cite{Lu14prl}. }
\label{fig:WL-WAL}
\end{figure}

In two dimensions, the conductivity correction from the quantum interference takes the form of logarithmic function with $\ell_\phi$ and $\ell$ serving as two cutoffs
\begin{eqnarray}
\sigma^{qi} \propto \pm \frac{e^2}{\pi h} \ln \frac{\ell_\phi}{\ell}
\end{eqnarray}
where $e^2/h$ is the conductance quantum, $-$ corresponds to weak localization and $+$ to weak anti-localization. Remind that $\ell$ is determined by the elastic scattering so it does not depend on temperature, while $\ell_\phi$ is determined by the inelastic scattering so it is usually a function of temperature, and empirically\cite{Thouless77prl}, $\ell_\phi \propto  T^{-p/2}$,
where $p$ is positive and depends on dephasing mechanisms and dimensionality\cite{Lee85rmp}.

When lowering temperature, $\ell_\phi$ could be longer and $\sigma^{qi}$ will become prominent, this gives the temperature dependence of the conductivity as shown in Fig. \ref{fig:WL-WAL} (b). Remember that $\sigma^{qi}$ is from the quantum interference between time reversed scattering loops [see Fig. \ref{fig:transport} (c)], so by applying a magnetic field that breaks time-reversal symmetry, $\sigma^{qi}$ can be destroyed, giving rise to the magnetoconductivity [see Fig. \ref{fig:WL-WAL} (a)].

\subsection{Weak anti-localization in topological insulators}

\begin{table}[htb]
  \centering
      \caption{The earlier experiments that observed the weak anti-localization in topological insulators. The magnetoconductivity measured was fitted by the Hikami-Larkin-Nagaoka formula\cite{HLN80} with two fitting parameters: the prefactor $\alpha$ and phase coherence length $
      \ell_\phi$. $^*$Calculated from $B_{\phi}\equiv\hbar/(4e\ell_\phi^2)$. Adapted from Ref. \cite{Lu11prb}.}
\label{tab:WAL-exp}
  \centering
\begin{tabular}{cccccc}
    \hline\hline
   Ref. & Sample & Thickness & $T$ (K)& $\alpha$  & $\ell_\phi$ (nm) \\ \hline
 \cite{Checkelsky09prl,Checkelsky11prl} &  Bi$_2$Se$_3$ & 10$\sim$20 nm &  0.3 & -0.38 &  \\ \hline
 \cite{Peng10natmat} &  Bi$_2$Se$_3$ & 10 nm  &  1.8 & -0.5$\sim$-0.38 & 106$\sim$237  \\ \hline
 \cite{Chen10prl} &  Bi$_2$Se$_3$ & 10 nm  &  1.8 & -0.5$\sim$-0.38 & 106$\sim$237$^*$  \\ \hline
 \cite{He11prl} &     Bi$_2$Te$_3$ & 50 nm &  2 & -0.39 & 331 \\ \hline
 \cite{Liu11prb} &     Bi$_2$Se$_3$ & 2$\sim$6 QL &  1.5& -0.39 & 75$\sim$200 \\ \hline
 \cite{Wang11prb} &     Bi$_2$Se$_3$   & 45 nm &   0.5 & -0.31  & 1100  \\
 &    (Bi,Pb)Se$_3$  &  &   & -0.35 & 640 \\ \hline
 \cite{Chen11rc} &  Bi$_2$Se$_3$ & 5$\sim$20 nm  &  0.01$\sim$2 & -1.1$\sim$-0.4 & 143$\sim\infty$  \\ \hline
  \cite{Kim11prb} &     Bi$_2$Se$_3$ & 3$\sim$100 QL & 1.5 & -0.63$\sim$-0.13 & 100$\sim$1000  \\\hline
  \cite{Steinberg11prb} &     Bi$_2$Se$_3$ & 20 nm & 0.3$\sim$100 & -1.1$\sim$-0.7 & 15$\sim$300  \\ \hline
  \end{tabular}
\end{table}

Since the studies of the carbon nanotube and graphene, it has been known that a two-dimensional gapless Dirac cone could have the weak anti-localization\cite{Suzuura02prl,McCann06prl}.
The surface states of a three-dimensional topological insulator are also two-dimensional gapless Dirac fermions. The weak anti-localization was observed soon after the discovery of Bi$_2$Se$_3$ and Bi$_2$Te$_3$ as topological insulators (see Table \ref{tab:WAL-exp}). There is another reason that the weak anti-localization was observed easily. Because of poor sample quality, the mean free path is short in the materials, of the order of 10 nm. But the phase coherence length can reach up to 100nm to 1$\mu$m below the liquid helium temperature. In other words, these materials are in the quantum diffusion regime at low temperatures, where the weak (anti-)localization are supposed to occur.

In the experiments, the magnetoconductivity was fitted by the Hikami-Larkin-Nagaoga formula for conventional electrons,\cite{HLN80}
with two fitting parameters: $\ell_\phi$ the phase coherence length and $\alpha$ a prefactor of the order of 1. $\alpha$ is positive for weak localization and negative for weak anti-localization.
In the experiments, $\alpha$ covers a wide range between around -0.4 and -1.1, suggesting that the observed WAL can be interpreted by considering only one or two surface bands,\cite{Checkelsky09prl,Peng10natmat,Checkelsky11prl,Chen10prl,He11prl,Liu11prb,Wang11prb,Chen11rc,Kim11prb,Steinberg11prb}
despite of the coexistence of multiple carrier channels from bulk and surface bands at the Fermi surface.
Even, the sharp WAL cusp can be completely suppressed by doping magnetic impurities only on the top surface of a topological insulator.\cite{He11prl}

\section{WEAK ANTI-LOCALIZATION AND LOCALIZATION CROSSOVER}
\label{sec:crossover}

\subsection{Theoretical formalism}

\begin{figure}[tbph]
\centering
\includegraphics[width=0.52\textwidth]{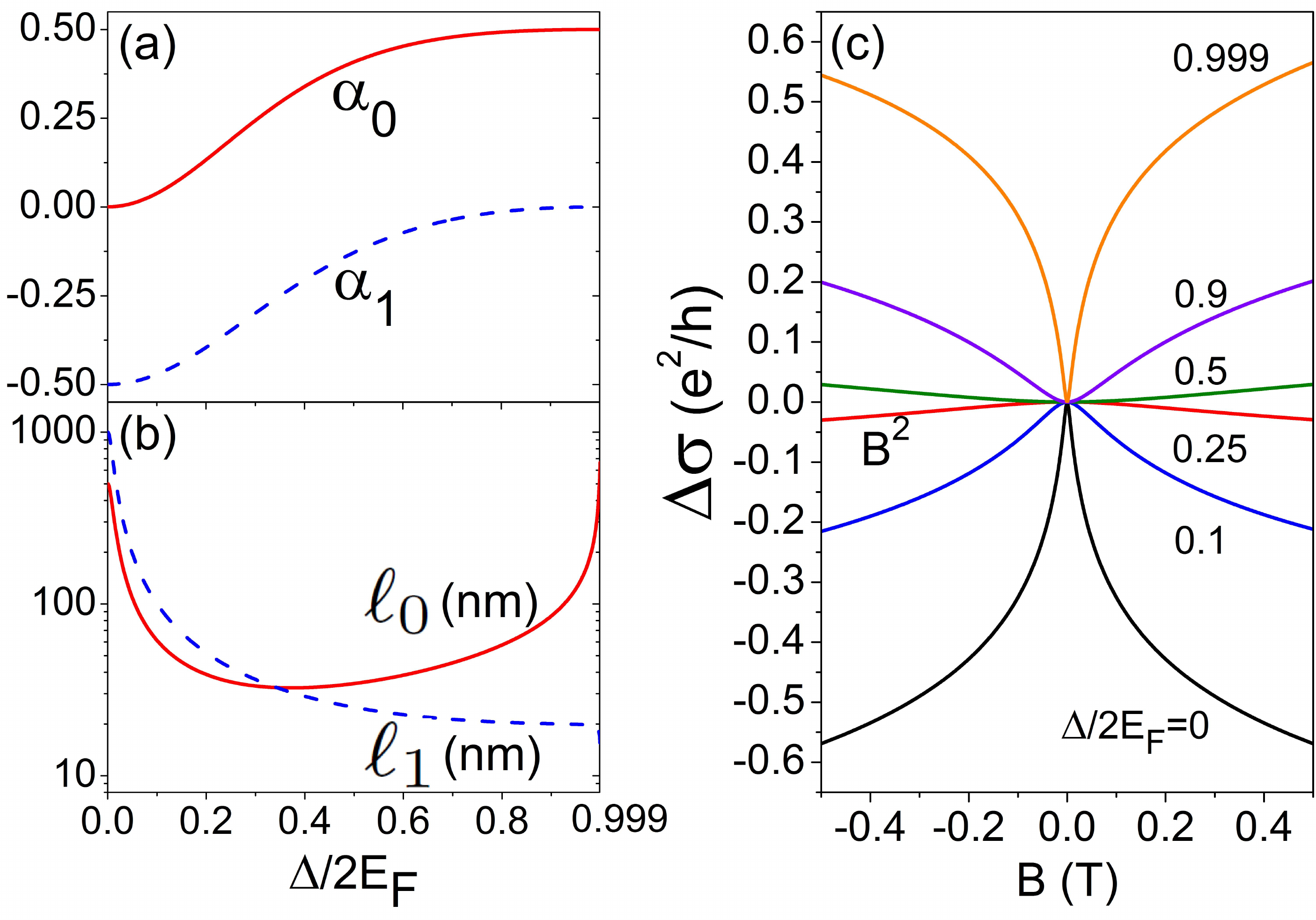}
\includegraphics[width=0.3\textwidth]{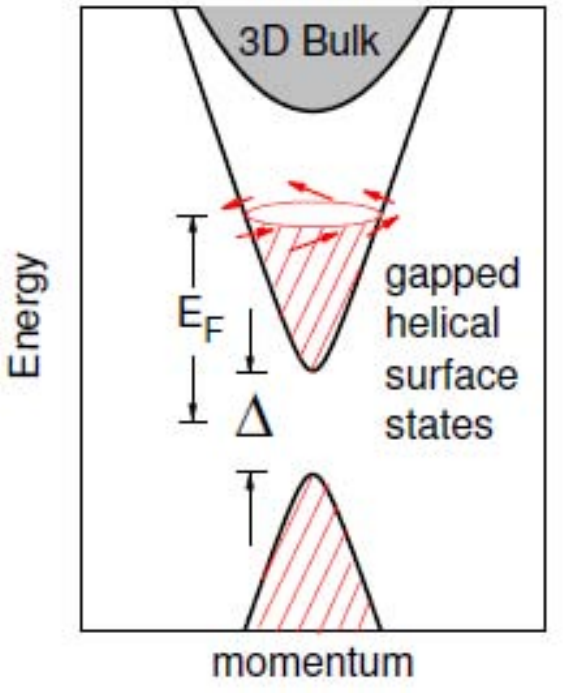}
\caption{(a) WL ($\protect\alpha_0$) and WAL ($\protect\alpha_1$) weight
factors as functions of $\Delta/2E_F$, where $\Delta$ is the gap, $E_F$ is
the Fermi energy. (b) WL ($\ell_0$) and WAL ($\ell_1$) lengths as functions
of $\Delta/2E_F$. (c) Magnetoconductivity $\Delta\protect\sigma(B)$ for
different $\Delta/2E_F$ in the limit of weak magnetic scattering. (Right) The gapped surface states as massive Dirac fermions. $\ell_{\protect\phi}=300$ nm. Adapted from Ref. \cite{Lu11prl}.}
\label{fig:mc_gap}
\end{figure}

The Dirac model in two dimensions is given by\cite{Shen12book,Shen11SPIN},
\begin{equation}\label{model}
H=\gamma (\sigma\times \mathbf{k})\cdot \hat{z}+\frac{\Delta}{2}\sigma_z,
\end{equation}
where $\gamma=v\hbar$, $v$ is the effective velocity, $\hbar$ is the reduced Planck constant, $\sigma=(\sigma_x,\sigma_y)$ are the Pauli matrices, and $\mathbf{k}=(k_x,k_y)$ is the wave vector. $H$ describes one conduction and one valence band, separated by a gap $\Delta$ [see Fig. \ref{fig:mc_gap} (right)]. We assume that the Fermi energy $E_{F}$ crosses the conduction band. The spinor wave function of the conduction band of $H$ is given by $\psi_{\mathbf{k}}(\mathbf{r})=(\cos\theta/2,-i\sin\theta/2e^{i\varphi})^Te^{i\mathbf{k}\cdot \mathbf{r}}$, with $\cos\theta\equiv \Delta/2E_F$ and $\tan\varphi=k_y/k_x$.

The model can be applied to several different cases. (1) The gapless surface electrons with $\Delta=0$.\cite{Shan10njp} (2) The massive surface states for a magnetically doped topological insulator with $\Delta \neq 0$.\cite{Chen10science,Wray11np} (3) The surface electrons in topological insulator thin films, where a finite gap is opened due to the finite size effect\cite{Lu10prb}. (4) The bulk electrons in topological insulator thin films where the mass term is the band gap between the conduction and valence band.\cite{Lu11prb}

The Berry phase is a geometric phase collected in an adiabatic cyclic process.\cite{Pancharatnam56,Berry84} The time-reversed scattering loops in Fig. \ref{fig:transport} (c) are equivalent to moving an electron on the Fermi surface by one cycle. As a result, the electron picks up a Berry phase
\cite{Lu11prl}
\begin{eqnarray}
\phi_b\equiv -i \int_0^{2\pi} d\varphi \left\langle \psi_\mathbf{k}(\mathbf{r})\left|\frac{\partial}{\partial \varphi}\right|\psi_{\mathbf{k}}(\mathbf{r})\right\rangle =\pi\left(1-\frac{\Delta}{2E_F}\right).
\end{eqnarray}
The Berry phase can give an explanation to the weak anti-localization of two-dimensional Dirac fermions. In the massless limit, the Berry phase $\phi_b=\pi$, leading to a destructive quantum interference that suppresses the back scattering and enhances the conductivity, leading to the weak anti-localization \cite{Suzuura02prl,McCann06prl}. While in the large-mass limit, $\phi_b=0$, which changes the quantum interference from destructive to constructive, resulting in the crossover to the weak localization \cite{Lu11prl,Imura09prb,Ghaemi10prl}, as shown in Fig. \ref{fig:mc_gap}, as the Berry phase changes from $\pi$ to $2\pi$ or 0.

The WAL-WL crossover can be depicted by the Lu-Shi-Shen formula for the magnetoconductivity\cite{Lu11prl}
\begin{equation}\label{MC}
\Delta \sigma (B)=\sum_{i=0,1}\frac{\alpha _{i}e^{2}}{\pi h}
\left[ \Psi (\frac{\ell_{B}^{2}}{\ell _{\phi i}^{2}}+\frac{1}{2})-\ln (\frac{\ell _{B}^{2}}{\ell_{\phi i }^{2}})\right] ,
\end{equation}
where $\Psi $ is the digamma function, $\ell _{B}^{2}\equiv \hbar/(4e|B|) $ is the magnetic length, $1/\ell_{\phi i}^2\equiv 1/\ell_{\phi}^2+1/\ell^2_i$, $\ell_{\phi}$ is the phase coherence length. The formula has two terms, one is for weak localization and the other is for weak anti-localization, characterized by two weight factors $\alpha_0\in[0,1/2)$ and $\alpha_1\in[-1/2,0)$, respectively. Sometimes $\alpha_i$ and $\ell_i$ were used by experiments as fitting parameters, but they are functions of $\Delta/2E_F$ [see Figs. \ref{fig:mc_gap}(a) and (b)], so the Lu-Shi-Shen formula has only two fitting parameters $\Delta/2E_F$ and $\ell_\phi$.

\subsection{Experimental verification}

The crossover from weak anti-localization to weak localization was soon observed by many experiments, where the gap is opened mainly by two approaches: (1) Magnetically doped surface states. The surface states of topological insulator Bi$_2$Se$_3$ and Bi$_2$Te$_3$ are gapless because of time-reversal symmetry. By doping magnetic impurities that breaks time-reversal symmetry can open the gap, as has been observed by ARPES.\cite{Chen10science,Wray11np}
The WAL-WL crossover has been observed in magnetically doped topological insulators, such as 3-quintuple-layer Cr-doped Bi$_{2}$Se$_3$\cite{Liu12prl}, Mn-doped Bi$_2$Se$_3$\cite{Zhang12prb},
and EuS/Bi$_2$Se$_3$ bilayers.\cite{Yang13prb} After lowering the temperature below the Curie temperature, a crossover from WAL to WL is observed, indicating the relation between the ferromagnetism-induced gap opening and the crossover.
(2) The finite-size effect of the surface states. In an ultrathin film of topological insulator, two gapless Dirac cones at the top and bottom surfaces can hybridize to open a finite-size gap, transforming the gapless Dirac cones into two massive Dirac cones.\cite{Lu10prb} For a given ultrathin film, the finite-size gap is fixed. But remember that the crossover depends on $\Delta/E_F$, where $E_F$ can be tuned by the gate voltage. Therefore, a WAL-WL crossover as a function of the gate voltage is observed in a 4-quintuple-layer Bi$_{1.14}$Sb$_{0.86}$Te$_3$.\cite{Lang13nl}

\section{WEAK LOCALIZATION OF BULK STATES}
\label{sec:bulk}

\begin{figure}[tbph]
\centering \includegraphics[width=0.6\textwidth]{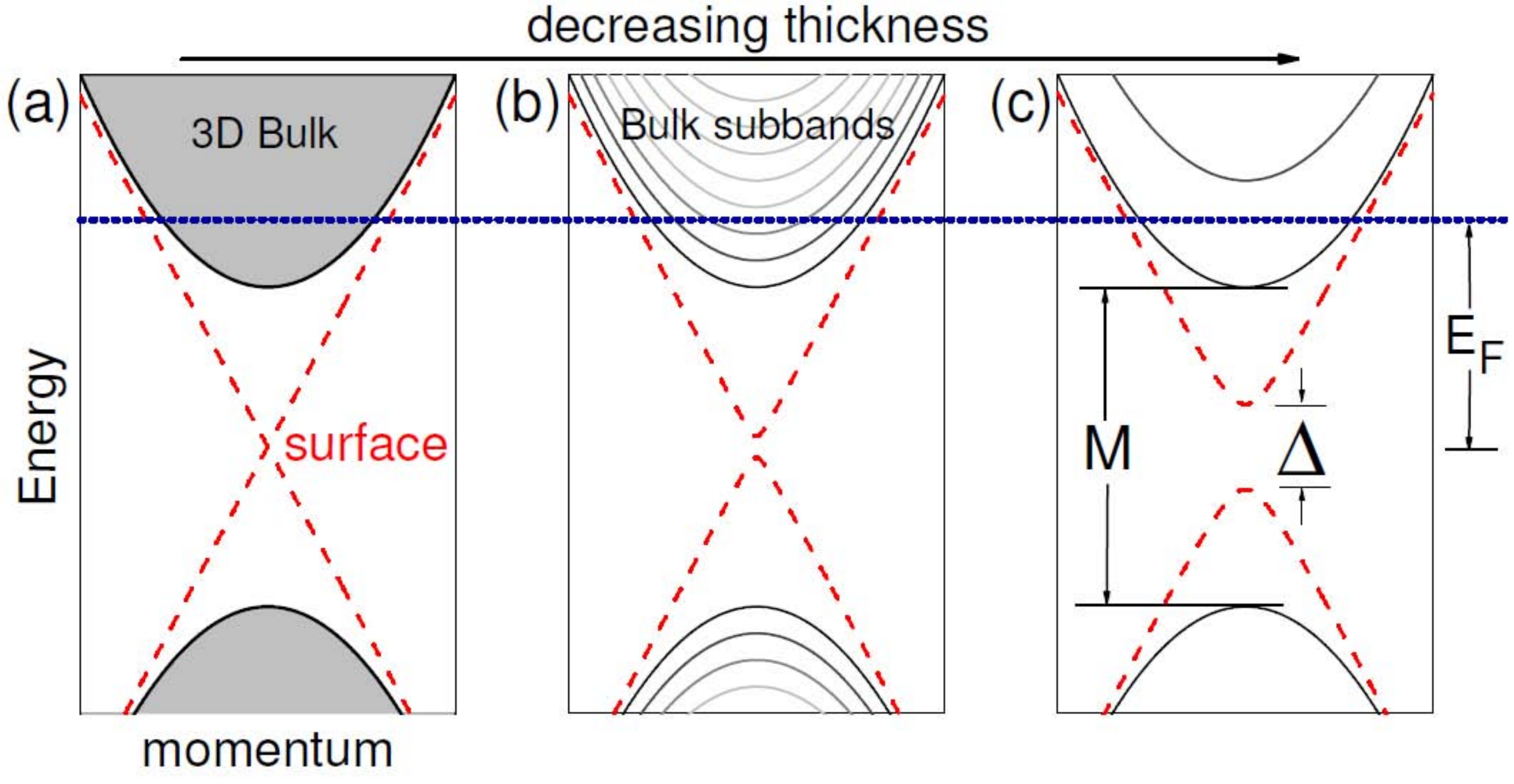}
\caption{(a) The gapped bulk (grey area) and gapless surface (dashed lines) bands of a 3D topological insulator.
(b) The quantum confinement along the $z$ direction splits the 3D bulk bands into 2D subbands, while the hybridization of the top and bottom surfaces opens a gap ($\Delta$) for the gapless surface bands. (c) In the ultrathin limit, the Fermi surface intersects with only one pair of bulk subbands (with band gap $M$) and one pair of gapped surface bands (each curve is two-fold degenerate). The horizontal dot line marks the Fermi energy $E_F$ measured from the Dirac point. Adapted from Ref. \cite{Lu11prb}.}
\label{fig:bulk}
\end{figure}

In most samples of topological insulator, the Fermi energy usually crosses not only the surface states, but also the bulk states [see Fig. \ref{fig:bulk}(a)]. The localization behavior of the bulk states was a controversial topic.

The minimal model to describe a 3D topological insulator is the modified Dirac model\cite{Shen11SPIN}
\begin{eqnarray}\label{H3D}
    H_{\mathrm{3D}}&=&
    \epsilon_{\mathbf{k}}+A\mathbf{k}\cdot \alpha+\mathcal{M}_{\mathbf{k}}\beta,
\end{eqnarray}
where $\mathbf{k}=(k_x,k_y,k_z)$ are wavevectors, the $4\times 4$ Dirac matrices can be expressed by the Pauli matrices $(\alpha_x,\alpha_y,\alpha_z)=\sigma_x\otimes (\sigma_x,\sigma_y,\sigma_z)$ and $\beta=\sigma_z\otimes\sigma_0$.
$\epsilon_{{\bf k}}=C+D(k_z^2+k^2)$, $\mathcal{M}_{{\bf k}}=m-B(k_z^2+k^2)$, $k_{\pm}=k_x\pm ik_y$, and $k^2=k_x^2+k_y^2$. $A$, $B$, $C$, $D$, and $m$ are model parameters. In a thin film of topological insulator, the 3D band structure split into 2D subbands. The simplest way to consider the lowest 2D bulk subbands in Figs. \ref{fig:bulk} (b) and (c) is to replace $\langle k_z\rangle=0$ and $\langle k_z^2\rangle=(\pi/d)^2$, where $d$ is the thickness of the film.
After defining $C+D (\pi/d)^2 =0$, $M/2\equiv m-B (\pi/d)^2$,
the Hamiltonian of the lowest 2D bulk subbands can be written as
\begin{eqnarray}\label{H3Dsub}
    H=D k^2+\tau_z(\frac{M}{2}-Bk^2)\sigma_z  + A (\sigma_xk_x+\sigma_yk_y).
\end{eqnarray}
with $\tau_z=\pm 1$ is the block index. This is nothing but a modified two-dimensional Dirac model, up to a unitary rotation in $x-y$ plane compared to Eq. (\ref{model}). At the band edge, the mass $M$ is large compared to the Fermi energy [Fig. \ref{fig:bulk} (c)]. As a result, we expect that weak localization happens for the bulk states\cite{Lu11prb} according to the Berry phase argument in the previous section. This result is quite different from other systems with strong spin-orbit interaction, where weak anti-localization is usually expected. This result was soon supported by other theoretical effort\cite{Garate12prb}. The crossover from weak anti-localization to localization can be realized by tuning the Fermi level or gate voltage in topological insulators. When the Fermi level is shifted from the bulk gap to the valence or conduction bands, there exist a competition between the  bulk electrons and surface electrons. In this case, it is possible to observe the weak localization. 

\section{COMPARISON WITH CONVENTIONAL ELECTRONS}
\label{sec:conventional}

\begin{table}[htbp]
\caption{Two-dimensional quantum diffusive transport of conventional and Dirac fermions in the presence of impurities of orthogonal (elastic), unitary (magnetic),
and symplectic (spin-orbit) symmetries.\cite{Dyson62jmp} Adapted from Ref. \cite{Shan12prb}.}
\label{tab:transport}%
\centering
\begin{tabular}{ccccc}
\hline
&   Scalar scattering &  Magnetic scattering   & Spin-orbit scattering  \\
  \hline
Conventional electron  &  WL & both suppressed & WL $\rightarrow$ WAL  \\
Massless Dirac fermion & WAL & suppressed WAL & WAL   \\
Dirac fermion in large-mass limit  & WL & suppressed WL &  suppressed WL    \\ \hline
\end{tabular}
\end{table}

Either the weak localization or weak anti-localization has been observed in conventional two dimensional electron systems\cite{Bergmann84PhysRep}. In conventional systems with dispersion of $p^2/2m$ and two-fold spin degeneracy on the Fermi surface, with only scalar disorder scattering, there will be weak localization. But strong spin-orbit scattering can leads to a crossover to the weak anti-localization,\cite{HLN80} because the spin-orbit scattering can bring the symplectic symmetry. For 2D gapless Dirac fermions, the symplectic symmetry is given by the band structure, so even with only scalar disorder scattering, one can still expect weak anti-localization. By putting extra spin-orbit scattering will not change the symmetry, so there is still weak anti-localization. The difference is in the large-mass limit of the massive Dirac fermions, where the weak localization is sensitive to the spin-orbit scattering and will be suppressed.

\begin{table}[htbp]
\caption{``Cooperon" channels for the conventional electrons and Dirac fermions.
Triplet (singlet) channel gives WL (WAL).
Spin-orbit scattering only quenches the triplet channels, leading to the crossover from WL to WAL for conventional electrons and the suppression of WL in the large-mass limit of Dirac fermions. Adapted from Ref. \cite{Shan12prb}. }
\label{tab:cooperon}%
\centering
\begin{tabular}{ccc}
\hline
Cooperon channels&   ``Triplet"($\Rightarrow$WL) &  ``Singlet"($\Rightarrow$WAL) \\
\hline
Conventional electron&  $\times 3$ &  $\times 1$\\
Massless Dirac fermion   & $\times 0$ &  $\times 1$    \\
Large-mass Dirac fermion  &   $\times 1$ & $\times 0$ \\
\hline
\end{tabular}
\end{table}

A deeper understanding can be found in Table \ref{tab:cooperon}.
The theory of weak localization can be equivalent to the diffusive transport of the Cooperon, a quasiparticle formed by pairing the electronic states before and after the back scattering. For conventional electrons, the pairing of two spin $1/2$ electrons can form 3 triplet Cooperon channels with total angular momenta $j=1$ and one singlet Cooperon channel with $j=0$.\cite{Altshuler81jetp}
Each channel of the triplets gives the weak localization, while the singlet channel gives the weak anti-localization. Because the triplets outnumber the singlet, we usually have the weak localization. However, the spin-orbit scattering behaves like a spin-dependent magnetic scattering, it can destroy the triplet Cooperon channels because they have a nonzero angular momentum, while has no effect on the singlet Cooperon channel. As a result, in the presence of strong spin-orbit scattering,
one expects a crossover from weak localization to weak anti-localization.

Different from the conventional electrons, the massive Dirac model is a superposition of one singlet and one ``triplet",\cite{Shan12prb} in the massless and large-mass limits, respectively (see Table \ref{tab:cooperon}). As a result, the weak localization in the large-mass limit can be destroyed by the spin-orbit scattering.

\section{INTERACTION-INDUCED LOCALIZATION OF SURFACE ELECTRONS}
\label{sec:interaction}

\subsection{Transport paradox in measurements}

One of the merits of topological insulators was thought to be that their topologically protected surface states could not be localized.\cite{Hasan10rmp,Moore10nat} This is based on a property of two-dimensional massless and non-interacting Dirac fermions in the thermodynamic limit.\cite{Bardarson07prl,Nomura07prl} For a long time the cusp of negtaive magnetoconductivity serves as an precursor of this delocalization tendency.

However, although a negative magnetoconductivity as the signature of the weak anti-localization has been observed in most topological insulator samples, the temperature dependence of the conductivity does not indicate its corresponding enhanced conductivity with decreasing temperature (see Fig. \ref{fig:WL-WAL}) as expected for weak anti-localization.
Instead, in most experiments, a suppression of the conductivity with decreasing temperature is observed\cite{Wang11prb,Liu11prb,Chen11rc,Takagaki12prb,Chiu13prb,Roy13apl}, indicating a behavior of the weak localization, which is supposed to happen in ordinary disordered metals as a precursor of the Anderson localization\cite{Bergmann84PhysRep,Lee85rmp}.
In these samples, the bulk states coexist with the surface states, so the bulk states may contribute to the localization. However, if the surface states could not be localized, the total conductivity should saturate at a critical temperature, but this is not the experimental measurement. Thus experimental observation in magnetoconductance and finite temperature conductance presents a transport paradox in topological insulators.

\subsection{Theory of interaction effect}

\begin{figure}[htbp]
\centering
\includegraphics[width=0.7\textwidth]{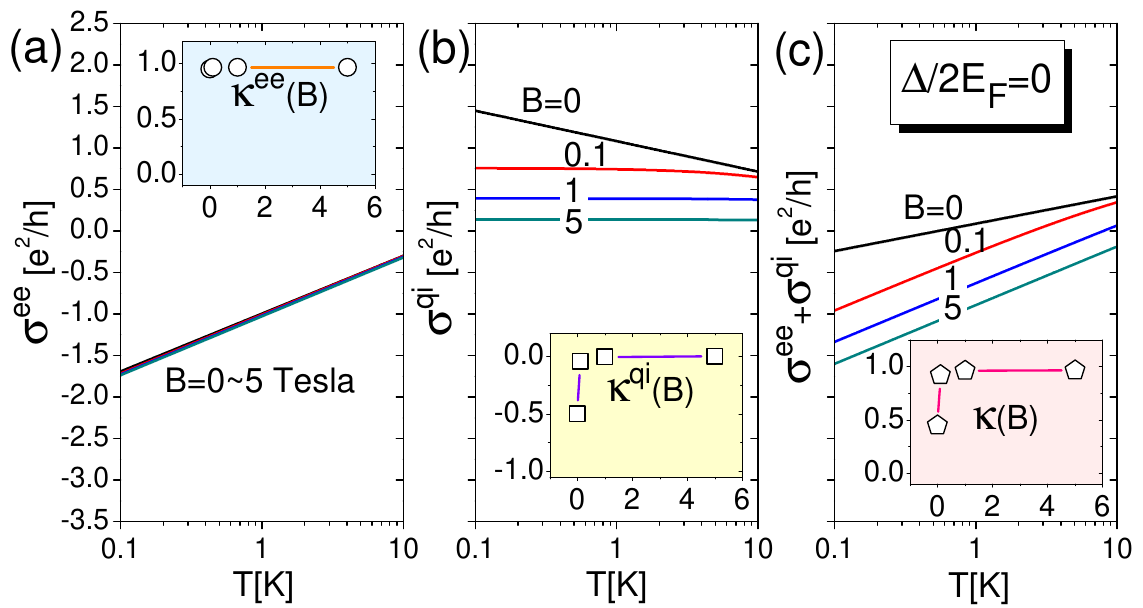}
\caption{The conductivity corrections from the electron-electron interaction ($\sigma^{ee}$) and quantum interference ($\sigma^{qi}$) as functions of temperature $T$ at different perpendicular magnetic fields $B$. Insets show the slopes $\kappa(B)\equiv(\pi h/e^{2})\partial\sigma/\partial\ln T$ at $T=1$ K as functions of $B$. The parameters are adopted from those in the topological insulators Bi$_{2}$Se$_{3}$ and Bi$_{2}$Te$_{3}$: $\Delta/2E_F=0$, $\gamma=3$ eV\AA, the relative permittivity\cite{Richter77psssb} $\varepsilon_{r}=100$, the mean free path $\ell=10$ nm, and the phase coherence length is taken to be $\ell_{\phi}=700\ T^{-p/2}$
nm with $T$ in units of Kelvin and $p=1$ \cite{Peng10natmat,Checkelsky11prl}. Adapted from Ref. \cite{Lu14prl}.}
\label{fig:conductivity}
\end{figure}

The transport paradox cannot be reconciled in the theory of the quantum interference correction ($\sigma^{qi}$) to the conductivity. The electron-electron interaction\cite{Altshuler80prl,Fukuyama80jpsj} ($\sigma^{ee}$) provides a possible way to understand the unexpected temperature dependence\cite{Lu14prl}. Figure \ref{fig:conductivity} compares $\sigma^{ee}$ and $\sigma^{qi}$ as functions of temperature at different magnetic fields, for the parameters comparable with those in the experiments.
With or without the magnetic field, $\sigma^{ee}$ decreases logarithmically with decreasing temperature [Figure \ref{fig:conductivity} (a)], showing a localization tendency, much like in usual two-dimensional disordered metals\cite{Altshuler80prl,Fukuyama80jpsj}. In contrast, zero-field $\sigma^{qi}$ enhances the conductivity when lowering temperature and is suppressed by the magnetic field [Figure \ref{fig:conductivity} (b)], a typical behavior of the weak anti-localization. The contribution from $\sigma^{ee}$ is stronger, so the overall temperature dependence of the conductivity shows a weak localization tendency [Figure \ref{fig:conductivity} (c)], consistent with the experimental observations.

\begin{table}[htbp]
\caption{The slope $\kappa\equiv(\pi h/e^2)\partial \sigma/\partial \ln T$ in experiments. $\sigma$ is the conductivity. The change of slope $ \delta \kappa \equiv \kappa(B_c)-\kappa(0)$. Above a critical magnetic field, the slope saturates. $n$ is the sheet carrier density. $n$ for C and D are converted from their cubic carrier densities.}
\label{tab:experimental kappa}%
\centering
\begin{tabular}{rllllll}
\hline
Experiment   & A\cite{Chen11rc}  & B$_1$\cite{Wang11prb}  & B$_2$\cite{Wang11prb} & C\cite{Takagaki12prb} & D\cite{Chiu13prb}  &  E\cite{Roy13apl}\\
\hline
Compound & Bi$_2$Se$_3$ & Bi$_2$Se$_3$  & Pb$_x$Bi$_{2-x}$Se$_3$  & Cu$_x$Bi$_{2-x}$Se$_3$   & Bi$_2$Te$_3$ & Bi$_2$Te$_3$ \\
n [$10^{12}$/cm$^2$]  & 20e$^-$  & 102e$^-$  &  49.5e$^-$  & 6.72e$^-$  &  37h$^+$  &  120e$^-$ \\
Thickness [nm]  & 10   & 45  & 45  & 80  & 65  & 4    \\ \hline
 $ \kappa(B=0)$ & 0.68   &  0.86 &0.70  &  1.37  &  1.33  & 0.58\\
 $ \kappa(B=0.2)$ &  0.8   & -  & -   & -   &  -  & 0.89 \\
 $ \kappa(B=0.5)$ &  -  & -  &  -  &  -  &  -  & 0.98 \\
 $ \kappa(B=1)$ &   0.8  & -  &  -  &  -  &  -  & 1.04 \\
 $ \kappa(B=2)$ &   0.8  & 1.18  & 1.29   &  1.68  &  -  & 1.07 \\
 $ \kappa(B=3)$ &   0.8  & -  &  -  &  1.66  &  -  & - \\
 $ \kappa(B=5)$ &  0.8   & -  &  -  & 1.67 & 1.66  & 1.09  \\
 $ \delta \kappa $ & 0.12   & 0.32  & 0.58  &  0.30  & 0.33 & 0.51 \\
 \hline
\end{tabular}
\end{table}

The temperature dependence of the conductivity is characterized by the slope
\begin{eqnarray}\label{kappas}
\kappa\equiv \frac{\pi h}{e^{2}}\frac{\partial\sigma}{\partial\ln T}.
\end{eqnarray}
We summarize the data of $\kappa$ in several recent experiments (Tab. \ref{tab:experimental kappa}), where
$\kappa$ is always positive; and by applying a magnetic field, $\kappa$ increases then saturates after $B$ exceeds a saturation field $B_\phi$.
Compared with the experiments, the results presented in Fig. \ref{fig:conductivity} agree well with the experiments, with comparable change of the conductivity (several $e^2/h$), temperatures (0.1 to 10 K) and magnetic fields (0 to 5 Tesla) \cite{Wang11prb,Liu11prb,Chen11rc,Takagaki12prb,Chiu13prb,Roy13apl}.

In this way, we show that the two-dimensional massless Dirac fermions will intend to be localized when the electron-electron interaction  and disorder scattering are taken into account.

\section{SUMMARY}

In summary, we developed  a quantum transport theory for two-dimensional massive Dirac fermions by considering the disorder scattering up to the quantum interference correction and the interplay with electron-electron interaction. The theory has been successfully applied to account for transport measurements in the topological insulators, it (1) predicted the crossover between weak anti-localization and weak localization as the surface states of topological insulators acquires a mass in the case of magnetic doping or in the case of thin film; (2) shows that the bulk states in a topological insulator thin film could have weak localization, in contrast to other systems with strong spin-orbit interaction; (3) clarifies the Cooperon channels for the 2D massive Dirac fermions and their responses to disorders of different symmetries; (4) reconciles the experimentally observed transport paradox in topological insulators: coexistence of weak localization tendency in the temperature dependence of the conductivity and the weak anti-localization behavior in the magnetoconductivity.

The decreasing conductivity in $\ln T$ at low temperatures indicates the weak localization of the surface electrons in topological insulators. Theoretically the surface electrons ``inherits" topological properties from the bulk band structures. The bulk-surface correspondence manifests the existence of the surface electrons surrounding the boundary of a topological insulator. It was extensively accepted \cite{Moore10nat, Hasan10rmp,Qi11rmp} that when disorders and impurities are added at the surface, there will be scattering between these surface electrons, but they cannot become localized. The electron-electron interaction makes it different: the surface electrons can be localized at low temperatures in an interacting topological insulator. The topological classification of topological insulators and superconductors is based on the framework of random matrix theory, which is essentially a single-particle problem \cite{Schnyder08prb}. The theory is based on the symmetries in a single particle Hamiltonian, and predicts the existence of at most five classes of topological insulators and superconductors in each dimension. Thus the classification is limited to the non-interacting systems. Introduction of disorders and impurities still keeps the square matrix property of the Hamiltonian, and does not destroy the validily of topologiacl classification if the disorders and impurity does not break the symmetry. However, the electron-electron interaction leads to a many-body probelm. Thus the weak localization of surface electrons in a topological insulator brings us an important issue on the robustness of a topological phase in an interacting system, which deserves to be investigated furthermore.

\acknowledgments

This work was supported by Research Grant Council of Hong Kong under Grant No.: HKU 7051/11P.

\end{document}